\documentclass[conference]{IEEEtran}
\IEEEoverridecommandlockouts
\usepackage{cite}
\usepackage{amsmath,amssymb,amsfonts}
\usepackage{algorithmic}

\usepackage{graphicx}
\usepackage{url}

\usepackage[caption=false]{subfig}

\usepackage{bm}
\usepackage[square,sort,comma,numbers]{natbib}
\usepackage{textcomp}
\usepackage{xcolor}
\usepackage[colorinlistoftodos]{todonotes}
\def\BibTeX{{\rm B\kern-.05em{\sc i\kern-.025em b}\kern-.08em
    T\kern-.1667em\lower.7ex\hbox{E}\kern-.125emX}}
\begin{document}

\title{Towards Understanding of eSports Athletes' Potentialities: The Sensing System \\for Data Collection and Analysis}

\author{\IEEEauthorblockN{Alexander Korotin, Nikita Khromov, Anton Stepanov, Andrey Lange, Evgeny Burnaev, Andrey Somov*}
\IEEEauthorblockA{\textit{Center for Computational and Data-Intensive Science and Engineering (CDISE)} \\
\textit{Skolkovo Institute of Science and Technology}\\
Moscow, Russia \\
*a.somov@skoltech.ru}
}

\maketitle

\begin{abstract}
eSports is a developing multidisciplinary research area. At present, there is a lack of relevant data collected from real eSports athletes and lack of platforms which could be used for the data collection and further analysis. In this paper, we present a sensing system for enabling the data collection from professional athletes. Also, we report on the case study about collecting and analyzing the gaze data from Monolith professional eSports team specializing in Counter-Strike: Global Offensive (CS:GO) discipline. We perform a comparative study on assessing the gaze of amateur players and professional athletes. The results of our work are vital for ensuring eSports data collection and the following analysis in the scope of scouting or assessing the eSports players and athletes.
\end{abstract}

\begin{IEEEkeywords}
eSports, wearable sensing, pervasive computing, data collection, Internet of Things
\end{IEEEkeywords}

\section{Introduction}
Video gaming has steadily evolved from home entertainment towards the professional eSports industry within the past few years~\cite{esports-2018}. Indeed, eSports has been already recognized as a professional sport in many countries worldwide. Its global audience reached 380 million in 2018 and will reach more than 550 million in 2021~\cite{newzoo-2018}.

eSports is an organized gaming where the single players or teams compete against each other to achieve a specific goal by the end of the game. For the sake of understanding the player's performance and for conducting the analysis of game statistics the  professional players (or \textit{eSports athletes}) often rely on online services, e.g. CS:GO Demos Manager (https://csgo-demos-manager.com) or a team coach. Online game services provide personal  statistics while coaches perform the game analysis based on their experience rather than on science. 

In terms of research, there are three relevant research areas in eSports (i) affective computing~\cite{affective_comp}, (ii) prediction related research~\cite{prediction}, and (iii) social structures in teams~\cite{social_aspects}. To date, the research in these areas is fragmented and lacks the experimental work with eSports athletes for data collection, modelling and interpretation tasks which could play a prominent role in the evolution of eSports research and make it practically feasible.

The emergence of Internet of Things (IoT)~\cite{iot-2012} and 
pervasive data science~\cite{pervasive_datasc} paradigms enable us to instrument the physical environment with myriads of sensors and create an interconnected world generating huge volumes of heterogeneous data. IoT allows for connecting heterogeneous and autonomous sensors~\cite{poco}, systems and services in a global network enriched with intellectual capabilities~\cite{Sasidharan}. It makes all 'things' in the scope of IoT available any time from anywhere. 
 
Data is getting a central tenant in computing applications where data collection, processing, and inference are of vital importance. Most of these ideas have been realized and deployed in the numerous smart-x applications including smart cities, smart transport, smart environment~\cite{iot-apps} and many others. These application are based on pervasive data science which focuses on data in pursuit of the concept of ubiquitous computing.

In this work we report on the sensing system for eSports athletes which enables the collection and processing of heterogeneous data. The proposed system relies on the IoT paradigm: we use heterogeneous sensors and data, as well as involve a game service for additional data collection. Also, we demonstrate a scenario how we collect the data from eSports athletes and perform their analysis for the assessment of professional quality. 

This paper is organized as follows: in Section II we introduce the reader to the related works in the area, in Section III we present methodology used in this research. We describe experimental testbed and report on experimental results in Section IV. Finally, we provide concluding remarks and discuss our future work in Section V.

\section{Related Work}
\subsection{Affective Computing}
Since humans actively or subconsciously express affect in many different ways, there are also countless different approaches to detect signals of affect. They all begin with sensors that gather data about various aspects of the human subject. For example, a camera can be used to detect the facial expressions of a person or their posture, and a microphone can be used to record the person’s voice. The text they type could be another data source for affect recognition. There also exist many different sensors that record physiological data like the heart rate, the respiration rate, brain activity or sweating, to name a few. The choice of which of these sensor to use depends on factors such as the type of affect to detect, the desired availability of sensor data, and how much interference with the gameplay can be tolerated.  

Affective Computing usually relies on sensors. When using Affective Computing in games, a first important task is to assess if a certain sensing technology can be used for a certain game and to select the most suitable sensors. Since there is no such thing as an optimal sensing technology, these decisions need to be made for each individual game \cite{guthier2016affective}.

Eye tracking is a process for detecting gaze position, pupil size, saccades and  blinks \cite{bojko2013eye}.
Data analysis from eye-tracking studies has focused on synchronic indicators such as fixation (duration, number, etc.) or saccade (amplitude, velocity, etc.), rather than diachronic indicators (scanpaths or saliency maps) different metrics are available for comparing two scanpaths, using either distance-based methods (string edit technique or Mannan distance) or vector-based methods. Distance-based methods compare scanpaths only from their spatial characteristics, while vector-based approaches perform the comparison across different dimensions (frequency, time, etc.). These metrics are more or less complex and relevant depending on the situation to be analyzed \cite{le2013methods}. According to \cite{bradley2008pupil} pupil’s response during affective picture viewing reflects emotional arousal associated with increased sympathetic activity. There are also methods suggested in \cite{xu2011pupillary} and \cite{xu2018improving} to measure cognitive workload and fatigue using eye tracking devices.

Emotional facial expressions represent the facial displays of emotions which determine different patterns of muscular correlates, cognitive responses, and brain activation. Autonomic and central nervous systems cooperate in order to provide a coherent pattern of mimic responses to specific contextual cues \cite{balconi2012consciousness}. So there are computer vision methods to detect emotions. There are several emotion detection techniques  discussed in, for instance, \cite{nguyen2017efficient,rathee2016adaptive}. 

Heart Rate Variability (HRV) tends to decrease with an increase in mental effort, stress, and frustration. Using machine learning, these emotional states and workload changes can be identified automatically \cite{hansen1995icarus}. The Blood Volume Pulse (BVP) is linked to the heart rate and can be used for emotion or mental stress recognition\cite{guthier2016affective}.
The electrodermal activity, according to \cite{el2016game,tastan2011learning}, 
could indicate awareness, the level of trust, stress or cognitive load. 


\subsection{Metrics}
In \cite{el2016game,mellon2009applying} the authors described general terms and measures  for metrics and analysis. The most important terms for the  data are auditobility, entry event distribution, exit event distribution, virality, engagement and retention rate. There are several important game metrics. The first one are user metrics. These are metrics related to the people, or users, who play games, from the dual perspective of them: being either customers, i.e. sources of revenue, or players, who behave in a particular way when interacting with games. The second one is a performance metrics. These are metrics related to the performance of the technical and software-based infrastructure behind a game, notably relevant for online or persistent games. Process metrics are metrics related to the actual process of developing games.

\subsection{Gameplay input analysis}
The gameplay input data that can be received from several sensors is extremely important for describing the gamer manner of playing and the level of his experience. Along with high-level game events such as deaths, kills, jumps, weapon fires the direct raw input data from keyboard, mouse and other sensors can provide more prolific insights. The analysis of mouse movements and buttons pressings as well as keyboard usage behaviour can be considered as individual characteristics of each player, particularly for predicting his skill level \cite{Buckley2013predicting}.
However, it requires proficient feature engineering with the raw data. 
The mouse input data can be described in terms of features such as the mean and the STD of the player mouse path, the mean and the STD of the velocity, the mean of click duration (the time that the button was pressed) and the mean number of clicks per some period of time. Also, the distribution of click intensity over screen zones can be considered \cite{Kaminsky2008identifying}.
Mouse movements of $x$ and $y$ positions are useful features as well. Regarding keys features the duration the special control keys were pressed, the most frequently pressed keys and the combinations of keys, pressed together, were used \cite{Buckley2013predicting}. The gaze data from eye-tracker is also fruitful for analysis. It allows to classify the player skills and the patterns in player behavior \cite{Choi2015Eye,Choi2018Eye}.

\section{Data Collection Platform}

 \subsection{Platform Design}
\begin{figure}[!htb]
\includegraphics[width=\linewidth]{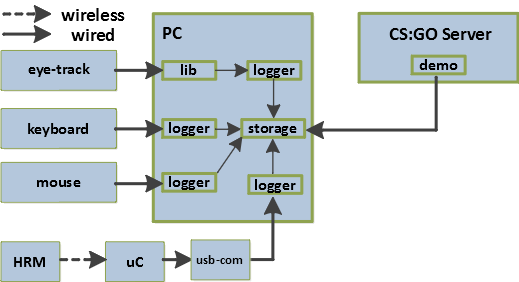}
\caption{Block diagram of data collection system.}
\label{block}
\end{figure}

 In this section we describe the proposed IoT platform for the eSports data collection and analysis.  The platform block diagram is shown in Figure~\ref{block}. It includes three units: \textit{sensing}, \textit{game PC} and \textit{server}, and allows the data collection from heterogeneous sensors and game service deployed on the CS:GO server. 
 
 In terms of \textit{sensing}, the proposed platform contains the following sensors: EyeLink eye-tracker, Garmin Heart Rate Monitor (HRM) belt, key/mouse logger, in-game data logger (CSGO HLTV demo). These sensors are used as they inherently allow for unobtrusive sensing of players and do not posses a high number of artefacts like, e.g. EEG. 
 
 We used the EyeLink eye-tracking software for eye-position measurements and developed a custom script for capturing the  data-stream from EyeLink library and recorded this data to the local \textit{csv} text file. We measure the gaze position 30 times per second.
 
The wireless HRM sensor transmits the data using the ANT ultra low-power protocol 4 times per second. This frequency allows us to measure the time of each heart beat since the pulse of a person is obviously less than 240 beats per minute. 
%
Upon receiving the data from the HRM sensor, we calculate the average number of heartbeats per minute and transmit this value to the PC using UART protocol. USB-UART adapter and virtual COM port, as well as the developed program, was used to receive the HRM data, get the current time and write the received data to a file.

Also, we developed a custom script for capturing mouse position and mouse/keyboard pressed keys. The data are collected every 10 ms and stored locally on the PC.

\begin{center}
\begin{figure}[!htb]
\includegraphics[width=\linewidth]{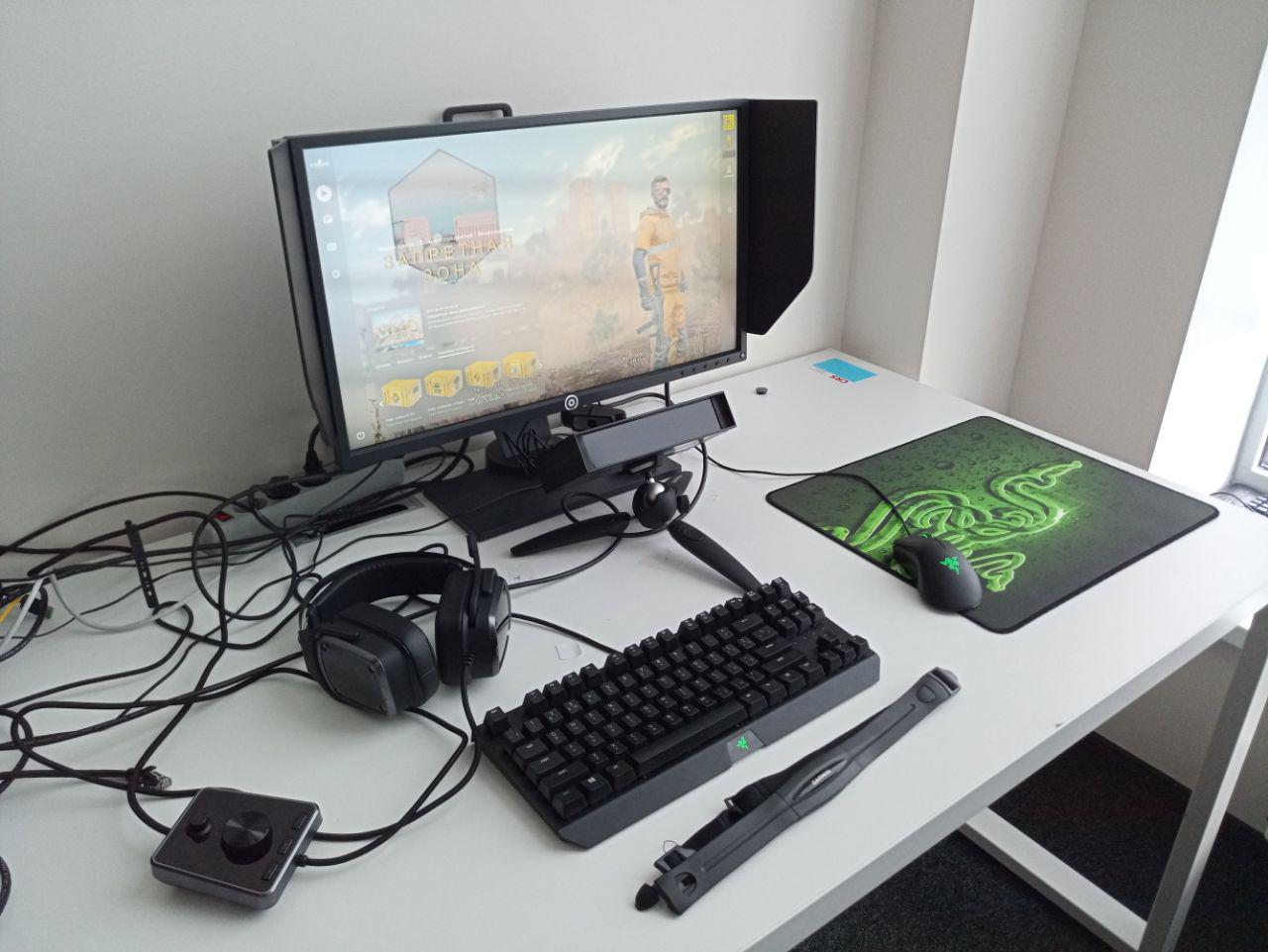}
\caption{Experimental testbed.}
\label{pic-sensor-system}
\end{figure}
\end{center}

 \textit{PC} unit is an advanced (storage, processing capability) gaming PC, where all external physiological sensors, i.e. eye tracker and HRM, are connected. Apart from the physiological data we perform logging of mouse and keyboard on the PC, as well as collection of game statistics. All required interfaces, libraries and loggers are installed on the game PC which acts as an experimental testbed (see Figure \ref{pic-sensor-system}).

The data from all the sources are stored on the PC except for the 'Demo' files which are recorded on a \textit{game server} and then are copied to the PC right after each game session. We record the in-game data during the game using the HLTV bot. 

\section{Methodology}
\subsection{Objectives}
It is essential for a coach and for every player interested in improving his skills to perform the analysis of the performance during both competition and training games. In most cases currently available tools provide analytics for the in-game data, e.g. a player accuracy based on the total number of weapon shots and the area of successful hits. Using wearable sensors we can assess a number of other metrics from game and evaluate the overall athlete 'quality'. The combination of those metrics could reveal weak and strong points of each athlete. The data has to be collected and preprocessed before the evaluation. 

\subsection{Game Scenario}
There are two roles for the Retake modification of CS:GO discipline: a terrorist team and counter-terrorist team. The terrorist team (2 players) is characterized by the defensive role. They have a bomb planted on the territory and have to defend it. The counter-terrorist team (3 players) has to deactivate the bomb or to kill the enemy team. The system shows the bomb location on the map at the beginning of each round. Players are also asked to buy exactly the same set of weapons for each round. Each round lasts for 40 seconds while there are 12 rounds in total. This scenario must be played without any breaks between the rounds. 

\subsection{Participants}
In this experiment players with different CS:GO experience were invited. Some players have not had CS:GO experience at all, while other players have spent impressive amount of time in this discipline ($>$10,000 hours), which means they play more than 4 hours per day. In total 15 players (5 professional and 10 amateur) participated in our experiment.

Prior to participate in the experiment, all participants were informed about the project, its goals, and the experiment. We obtained a written consent from each participant in the experiment. Afterwards, we ensured by a questionnaire that all the participants are in a good form and do not take any drugs in order to avoid the interference with the experimental results.

\section{Evaluation}
The platform we propose in this work allows for monitoring of the crucial parameters of an eSports athlete during the game. In this section we discuss our analysis in details.

\subsection{Gaze Analysis}
In this section, we provide the analysis of the gaze of professional athletes and amateur players. We state and discuss several important differences between the athletes and players.

\begin{center}
\begin{figure}[!htb]
\includegraphics[width=\linewidth]{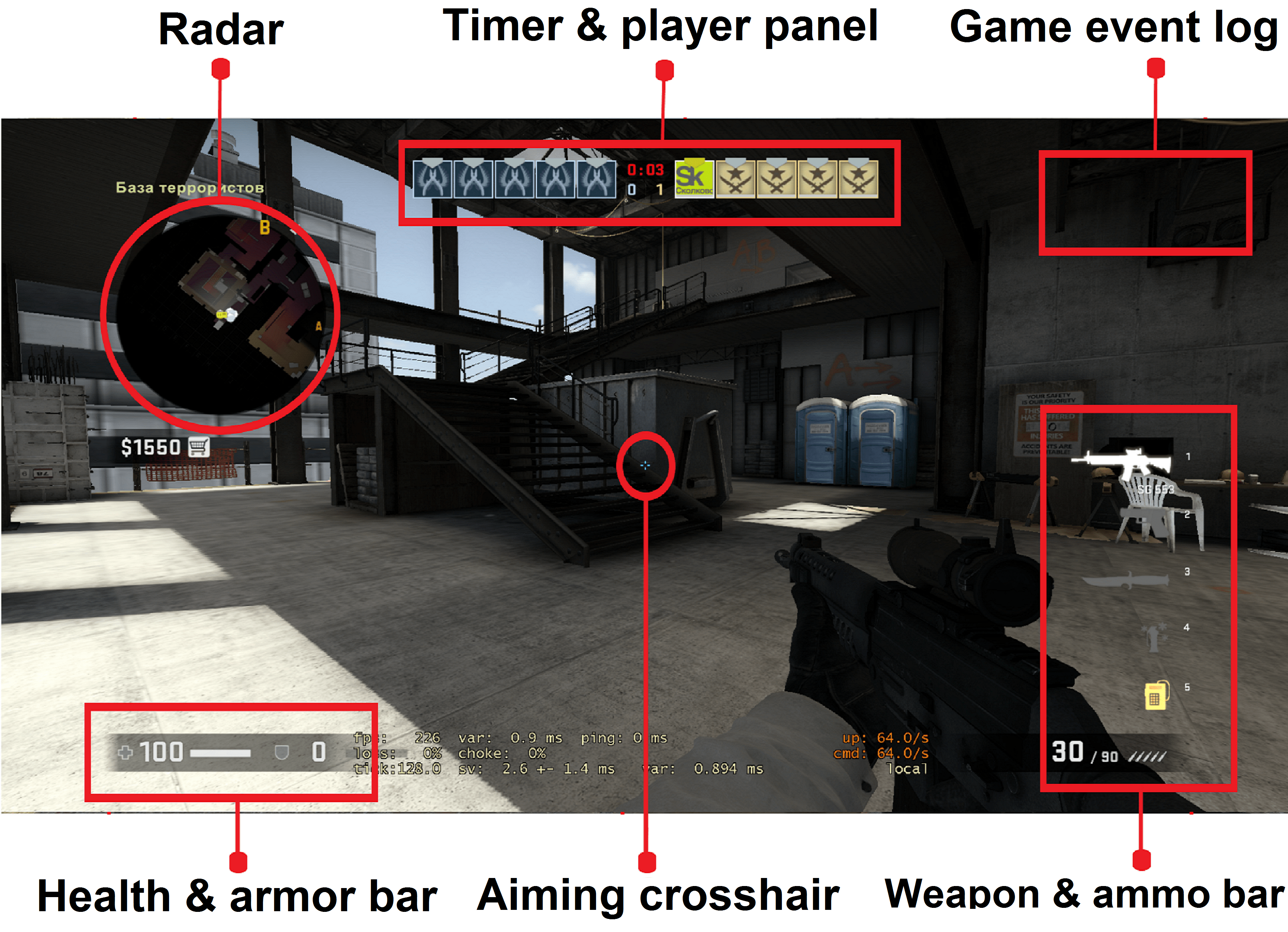}
\caption{Elements of the user interface in CS:GO.}
\label{figure:ui}
\end{figure}
\end{center}

During the Counter Strike game the cyber atheltes have to visually control several in-game elements. First, they have to track the dynamic 3D game environment (enemies, allies, shots, positions, etc.).  Second, the game player has to regularly check the game User Interface (UI) that overlay the game 3D environment. We show these UI elements in Figure \ref{figure:ui}.

The UI elements include the health \& armor panel, weapon \& ammo panel, in-game radar, kill \& death log, game timer \& player list. Controlling them is crucial indeed: the radar provides the player with the information about the teammates and enemy position, weapon \& ammo panel allows for efficient management of ammo usage, etc.

The gaze data set consists of 2-dimensional time series of moving screen coordinates $(x, y)$ for each of 15 players (5 professional and 10 amateur). Each time series is recorded at the frequency of 60 Hz and contains 10 repeated rounds lasting approximately 1 minute each. The gaze positions are within 1920x1080 screen size.

To pre-process the data we first extract the periods when the player was alive. 
To do this we used the parsed information from the game demo log file.
Next, we performed the linear interpolation of missing values of players' gaze. It was performed only for the gaps with missing data, that are less than $0.1$ seconds (6 missing consequent observations maximum). The total amount of missing data is about $4\%$ (situations when the eye-tracker looses player's gaze or player moves the gaze away from the monitor).

We denote the resulting prepossessed time series for the player $n=1,\dots,15$ by $(x_t^n, y_t^n)$,
where $t=1,\dots, t_{n}$ is the integer index of time. 
We assume that the gaze positions $(x, y)$ are mainly caused by the specific zone of interest shown on the screen during the game (see Figure \ref{figure:ui}).
By confirming this assumption one can then match the gaze position with a single categorical value representing such an area of interest, e.g. radar, weapon panel or aiming cross-hair.


We apply $k$-center clustering to the set of all observed gaze positions of all players $(x_{t}^{n}, y_{t}^{n})$ (for all $n, t$). We manually set $k=9$ cluster centers and use Euclidean metric as a distance. The cluster centers are given in Table \ref{table:cluster-centers} below.

\renewcommand{\figurename}{Table}
\begin{figure}[htb!]
\centering
\begin{tabular}{|l|c|c|l|}
\hline
\textbf{Zone} &  \textbf{x (center)} & \textbf{y (center)} & \textbf{Explanation} \\
\hline
No. 1 &         960 &         540 &      Aiming Cross-hair \\
No. 2 &         345 &         815 &             Radar Area \\
No. 3 &         310 &         180 &     Armor \& Health Bar \\
No. 4 &        1205 &         530 &    Right Area of Sight \\
No. 5 &        1610 &         180 &    Weapon \& Ammo Panel \\
No. 6 &         715 &         530 &     Left Area of Sight \\
No. 7 &        1575 &         815 &       Kill \& Death Log \\
No. 8 &         960 &         260 &   Bottom Area of Sight \\
No. 9 &         960 &         900 &  Timer \& Players Panel \\
\hline
\end{tabular}
\caption{Coordinates of the clusters' centers.}
\label{table:cluster-centers}
\end{figure}
\renewcommand{\figurename}{Fig.}
The obtained clustering in shown in Figure \ref{figure:clusters} below. Six of the clusters centers represent different UI-related zones of player's interest shown in Figure \ref{figure:ui}. Zone $1$ is the aiming cross hair, Zone $2$ is the radar area, Zone $3$ is the health \& armor panel, Zone 5 is the weapon \& ammo panel, Zone 7 is the kill \& death log, Zone 9 is the game timer \& player panel. We also add three zones around the aiming cross-hair that represent players gaze concentrated near the aiming cross-hair (right, left and below the cross-hair for Zones 4, 6 and 8 respectively).

\begin{center}
\begin{figure}[!htb]
\includegraphics[width=\linewidth]{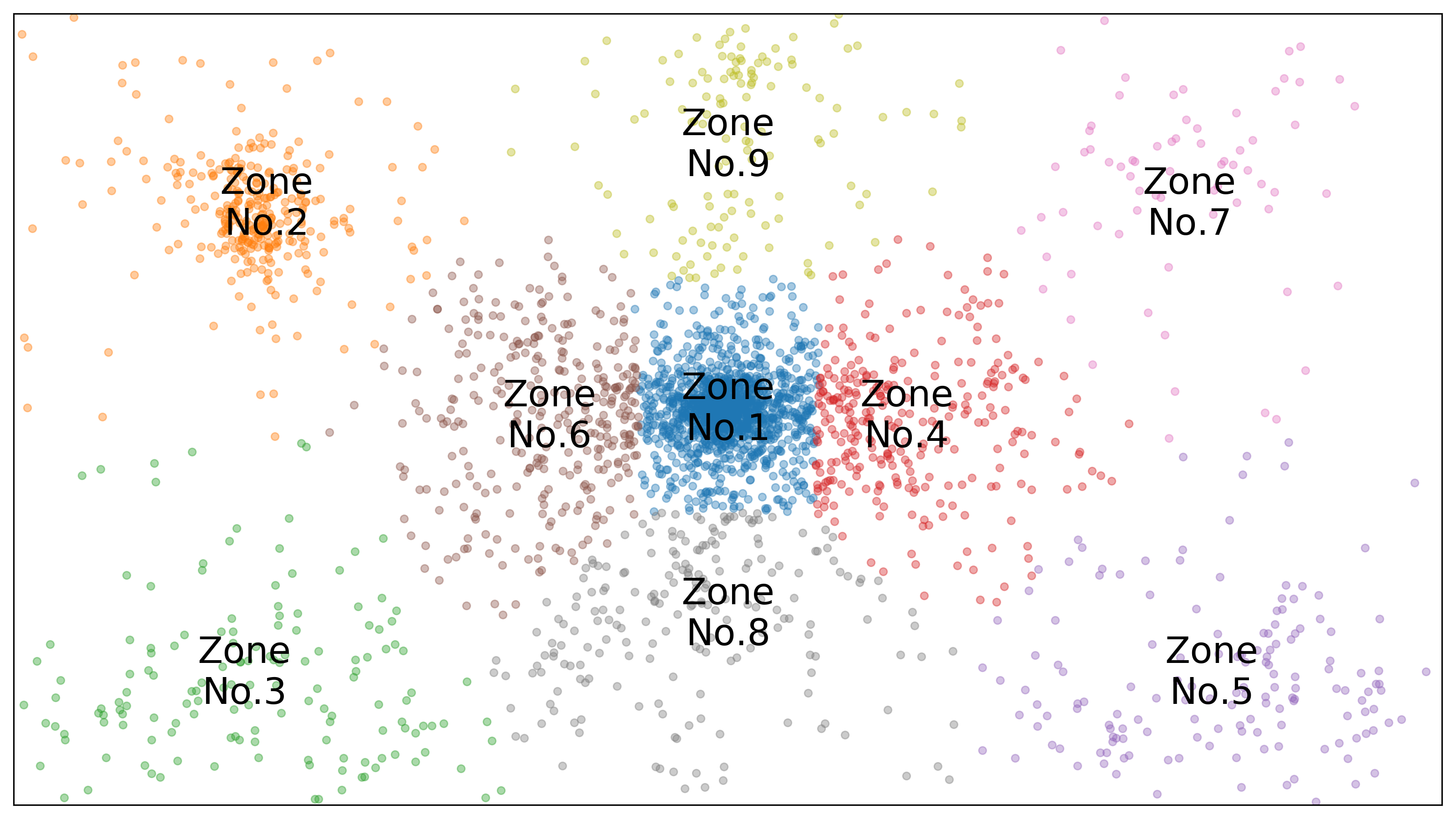}
\caption{Gaze positions (randomly shown 3000 of them) and the clusters shown by different colors. Some clusters correspond to a specific zone of interest.}
\label{figure:clusters}
\end{figure}
\end{center}

By considering 15s rolling window 
(indexed by ${\tau=1,\ldots,\tau_n}$) we calculate the probabilities over all clusters 
$p_{\tau, k}^n$, $\sum_{k=1}^K p_{\tau, k}^n=1$,
as the frequency when the gaze of $n$-th player was focused on the $k$-th zone of interest. Also, we consider similar distribution averaged over all rolling windows
\begin{equation}
    p_k^n = \frac{1}{\tau_n} 
    \sum_{\tau=1}^{\tau_n} p_{\tau, k}^n
    \label{equ:1}
\end{equation}
as a measure of comparative usage of the gaze zones for each player.

\begin{center}
\begin{figure}[!htb]
\includegraphics[width=\linewidth]{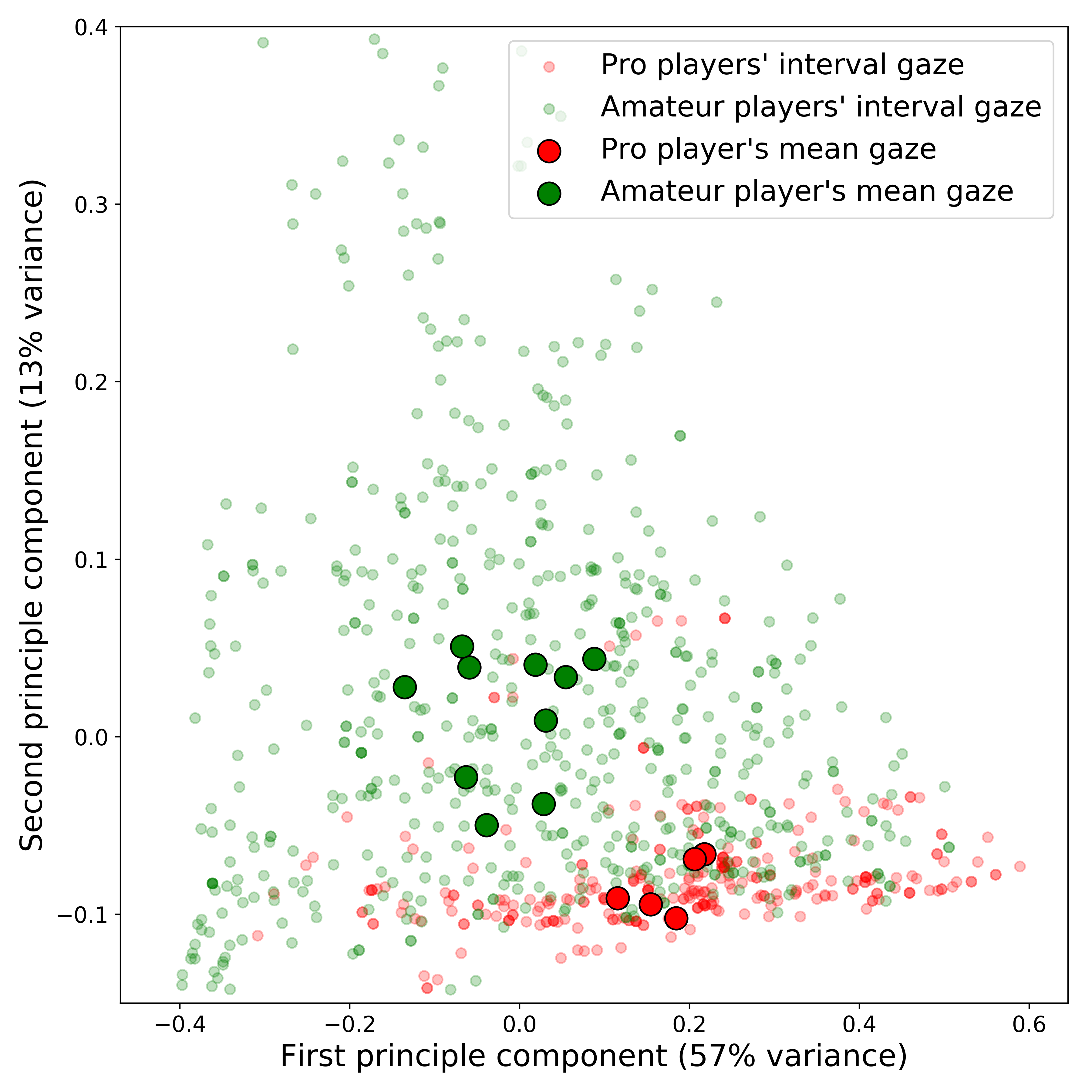}
\caption{First two principle components of players' gaze.}
\label{figure:main-components}
\end{figure}
\end{center}

We then raise the question whether there is a difference in the gaze of the professional athletes and amateur players and whether they can be clearly explained by simple rules. 
To answer this question we apply the Principal Component Analysis (PCA)
\cite{wold1987principal,jolliffe2011principal} 
to the set of all $K$-dimensional vectors of distributions
\begin{equation}
   [p_{\tau, 1}^n, \ldots, p_{\tau, K}^n], \quad \tau=1,\ldots, \tau_n.
   \label{equ:2}
\end{equation}
The projection of such data onto the linear subspace of the first two principle components (sorted by their variance) is shown in  Figure \ref{figure:main-components}.
Green points correspond to the gazes of amateur players and 
the red points correspond to the professional ones. 
Small points represent the projections of points (\ref{equ:2})
for some randomly chosen rolling windows $\tau$.
Large points are the projections of averaged distributions (\ref{equ:1}).

The first main component explains $57\%$ of the variance in the data, while the second component explains only $13\%$. Nevertheless we can clearly see that professional athletes can be easily separated from amateur players by these two components. To answer why this happens we provide a more detailed analysis.

Surprisingly, it turns out that both main components are almost sparse with only one dominant coordinate. The dominant coordinate in the first main component corresponds to the Zone $1$ (aiming cross-hair zone) while the dominant coordinate in the second main component corresponds to the Zone $2$ (radar zone) (see Figure \ref{figure:clusters}).

\begin{figure}[!htb]

\subfloat[Professional player's gaze.]{
  \includegraphics[clip,width=\linewidth]{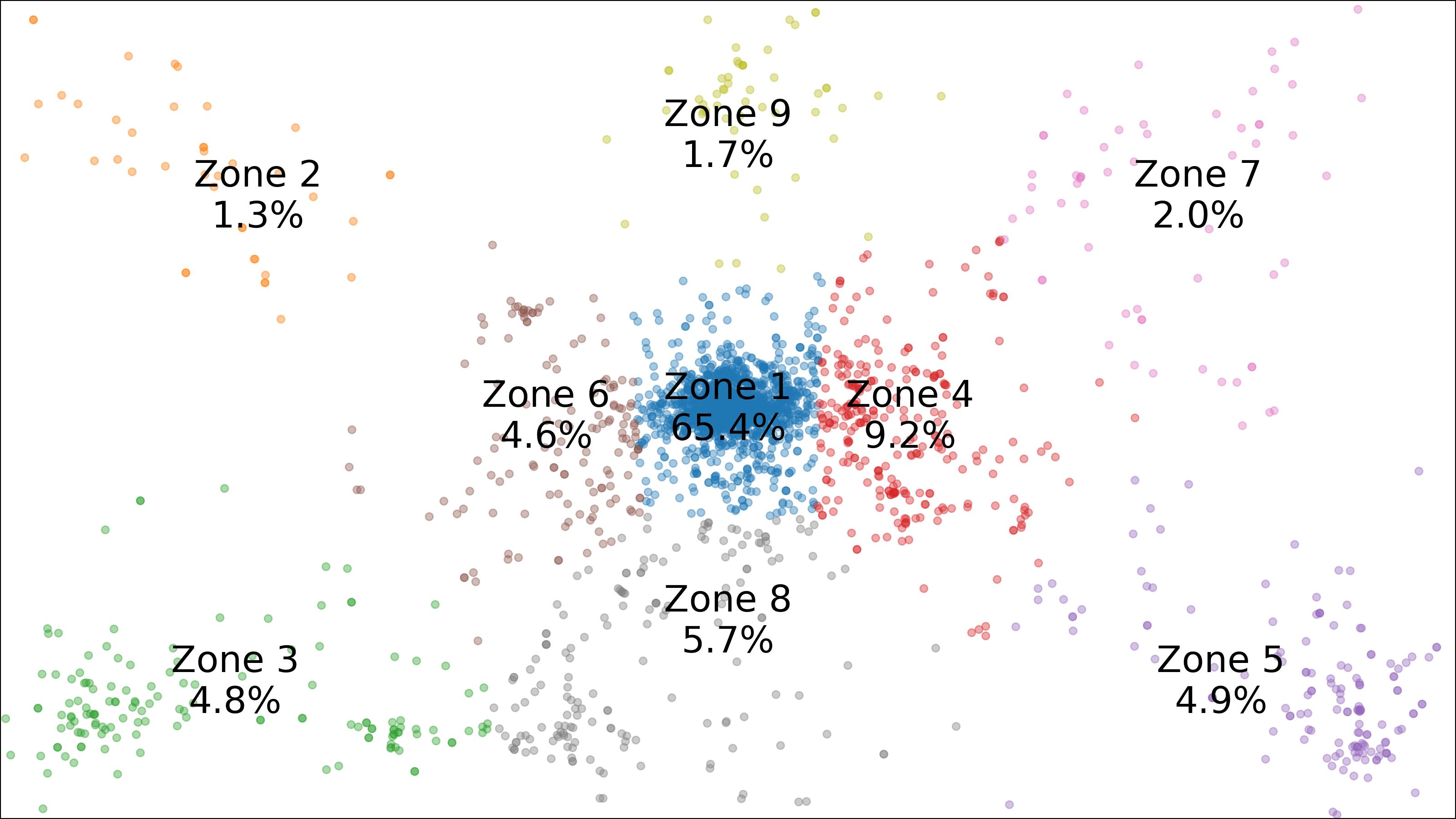}
  \label{figure:heatmap-pro}
}

\subfloat[Amateur player's gaze.]{
  \includegraphics[clip,width=\linewidth]{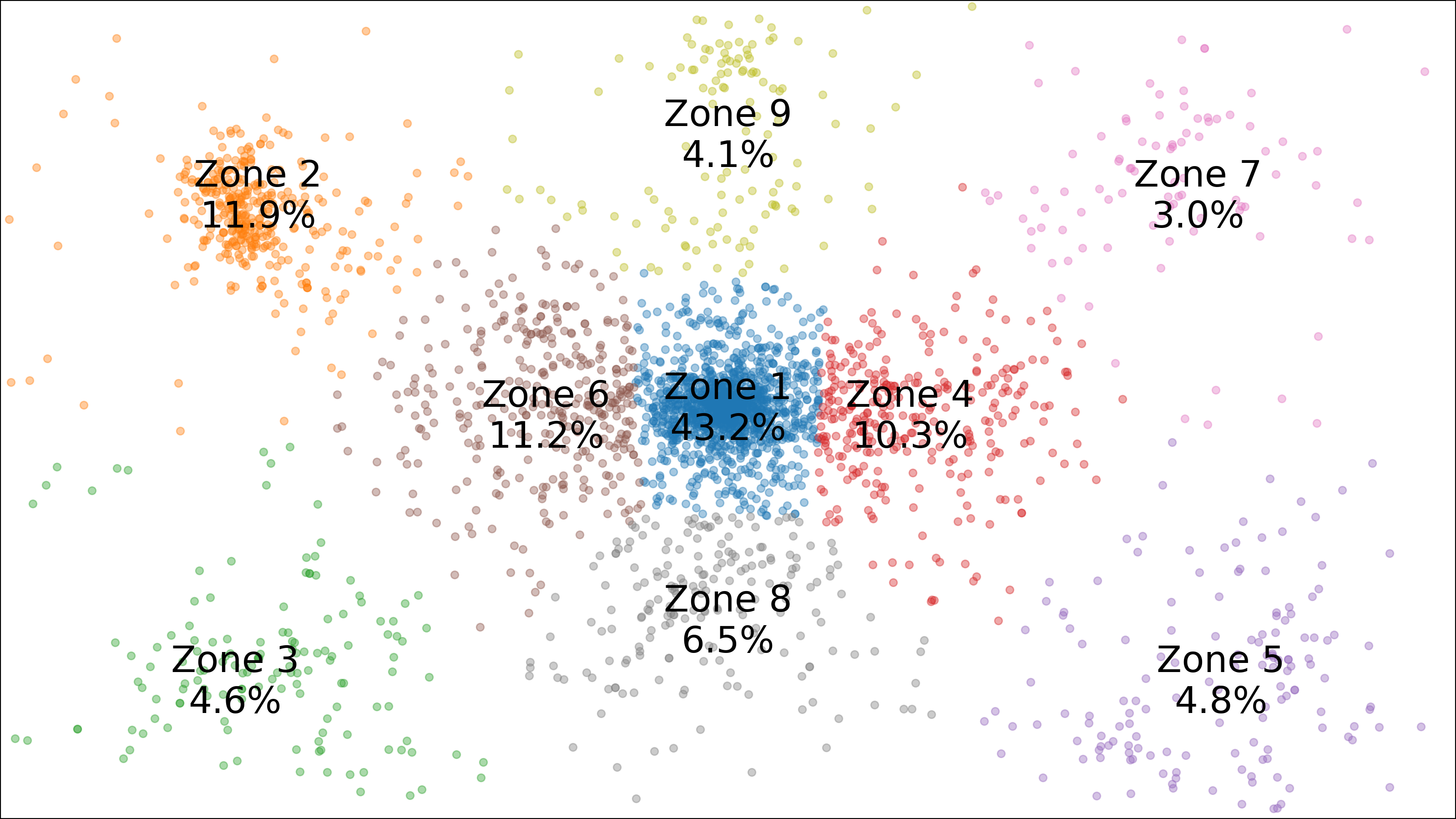}
  \label{figure:heatmap-amateur}
}
\label{figure:hedge-results}
\caption{Gaze heatmaps for the professional athletes and amateur players (3000 random gaze samples) with colored zones and percentage of all players' gaze concentration time on the zones.}
\end{figure}

Based on this analysis, we conclude that the professional athletes spend more time looking at the screen center than the  amateur players. At the same time athletes do not look at the radar too often as players do. We show the gaze heat map of an average professional athlete in Figure \ref{figure:heatmap-pro} and average amateur player in Figure \ref{figure:heatmap-amateur}. The gaze of the professional athlete is most of the time concentrated at the screen center while the amateur players spent more time looking at the UI elements, especially the game radar.

Since Counter Strike discipline is the first-person shooter game, it was reasonable to expect that players' gaze most of the time is concentrated around the screen center where the aiming cross-hair is located. The fact that the professional athletes spent more time looking at the aiming cross-hair is logical enough: they perfectly know the map and perfectly position themselves in the game --- that is why there is no need to look around. This also explains the point why they do not need to use radar frequently.

\subsection{Mouse \& Keyboard Usage Analysis}

In this section, we report on the analysis of the mouse and  keyboard usage by professional athletes and amateur players.

In the CS:GO discipline main control keyboard keys are common for most players:
\begin{enumerate}
    \item \textbf{W, S}  --- forward and backward movement keys;
    \item \textbf{A, D} --- left and right movement keys;
    \item \textbf{Mouse 1} --- weapon shooting key (left mouse button).
\end{enumerate}

The way how the player uses the control keys has an impact on a  player skill. For investigating this point we extract a number of features related to the usage of specific keyboard keys.
In this section, we discuss the features that highly correlate with the specific skill of players and athletes.

\begin{figure}[!htb]
\includegraphics[clip,width=\linewidth]{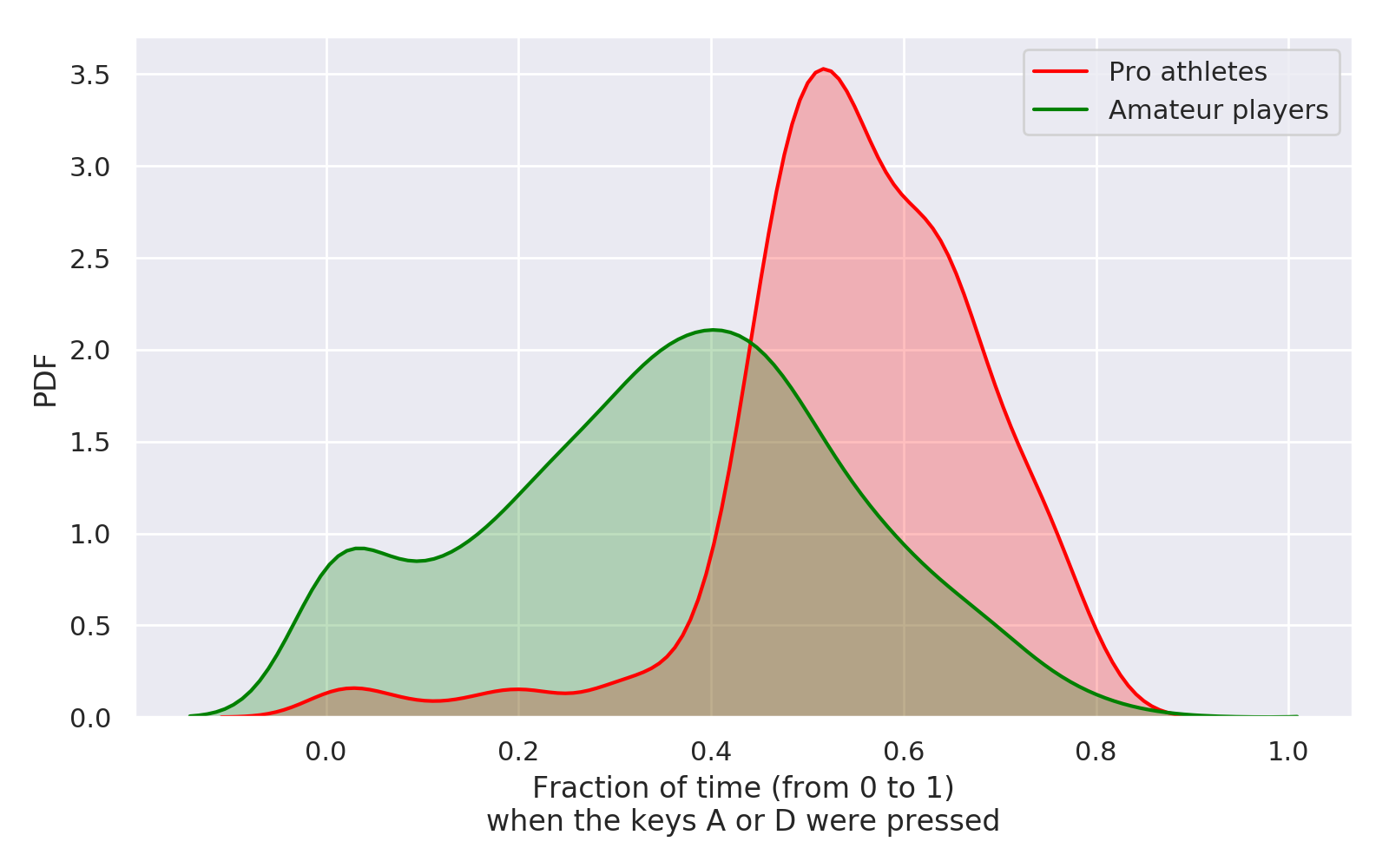}
\vspace{-7mm}\caption{Distribution of the usage of keys \textbf{A} or \textbf{D} by amateur players and professional athletes.}
\label{figure:ad-comparison}
\end{figure}

\begin{figure}[!htb]
\includegraphics[clip,width=\linewidth]{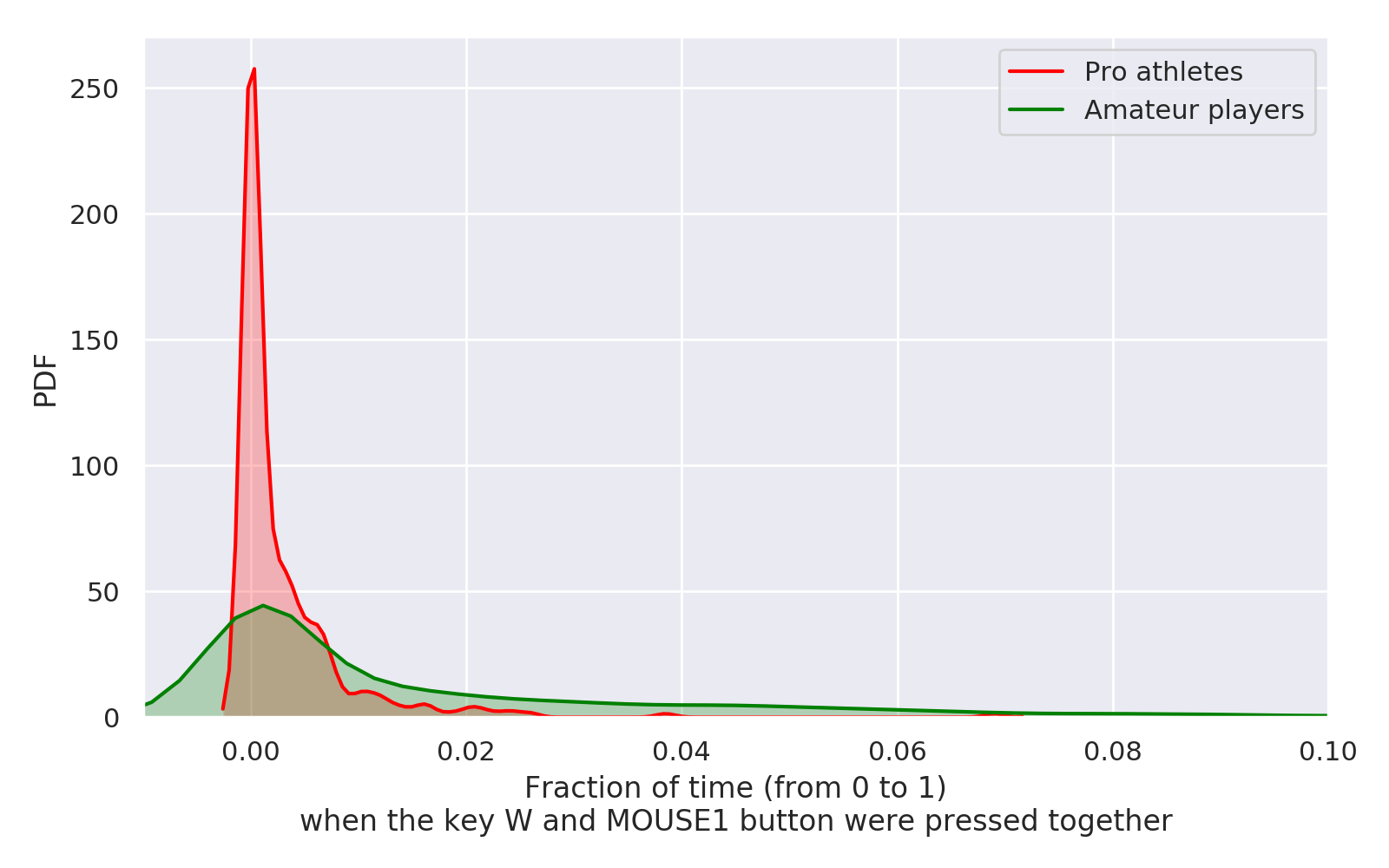}
\vspace{-7mm}\caption{Distribution of the simultaneous usage of \textbf{W} \& \textbf{MOUSE1} keys  by amateur players and professional athletes.}
\label{figure:wm1-comparison}
\end{figure}

First, we consider the period of time when the player used the \textbf{A} or \textbf{D} keys, i.e. movement to the right and to the left. Second, we examine the period of the time when the player pressed the shooting button \textbf{MOUSE1} while moving forward \textbf{W}. We plot the distribution of these features on Figure \ref{figure:ad-comparison} and Figure \ref{figure:wm1-comparison}, respectively. We note here that we smooth the plots by using Gaussian Kernel Density Estimation.

On the one hand, we can clearly see that the professional athletes often use the left and right movement keys, i.e. \textbf{A} and \textbf{D}. At the same time, amateur players often combine the moving forward (running) key with shooting. According to the game engine this action significantly reduces the shooting accuracy. This analysis confirms that professional athletes better control their movements in game scenario.

\vspace{-5mm}

\section{Conclusion}
In this work, we have demonstrated a platform for collecting the data from eSports athletes. The platform includes three types of sensors: physiological, environmental and game telemetry. Using this platform we carried out the experiment on the data collection from  professional athletes and amateur players. The special emphasis was made on the gaze analysis.

Our future research aims at the data collection from the  heterogeneous sensors and finding correlations between the  physiological and psycho-emotional conditions of eSports athletes. 

Analysis of eSports data requires the development of specialized machine learning methods: there are outliers due to faulty sensors, part of the data is represented in the form of multidimensional time series, and another part is provided as heterogeneous events streams, etc. Thus we should develop a unified methodology for eSports data processing, including pre-processing data to remove anomalies \cite{EnsemblesDetectors2015,ConformalAD2015,kNN2017}, methods of imbalanced classification \cite{Imbalance2019,Imbalance2015} and dimensionality reduction \cite{MLR2018,DRreg} to extract specific patterns and predict rare events.

\section*{Acknowledgment}
The reported study was funded by RFBR according to the research project No. 18-29-22077$\backslash$18.

Authors would like to thank Skoltech Cyberacademy, CS:GO Monolith team and their coach Rustam ``TsaGa'' Tsagolov for fruitful discussions while preparing this article. Also, the authors thank Alexey "ub1que" Polivanov for supporting the experiments by providing a slot at the CS:GO Online Retake server \footnote{\url{http://ub1que.ru}}.

\bibliographystyle{plain}
\bibliography{main}

\end{document}